\documentclass[prb, nofootinbib, twocolumn, amsmath, amssymb, superscriptaddress]{revtex4-2}

\usepackage{bbold}
\usepackage{hyperref}
\hypersetup{colorlinks=true, citecolor=blue, linkcolor=blue, urlcolor=gray}
\usepackage{graphicx}
\usepackage{bm}
\usepackage{color}
\usepackage{url}
\usepackage{tikz}
\usetikzlibrary{calc}
\usepackage{pgfplots}
\usepackage{amsmath,amssymb}
\usepackage{verbatim}
\usepackage{physics}
\usepackage{xcolor}
\usepackage{mwe}
\usepackage{tikz}
\usetikzlibrary{calc}
\usepackage{pgfplots}

\newcommand{\bea}{\begin{eqnarray}}
\newcommand{\eea}{\end{eqnarray}}

\newcommand{\be}{\begin{equation}}
\newcommand{\ee}{\end{equation}}

\newcommand{\beal}{\begin{align}}
\newcommand{\eeal}{\end{align}}

\newcommand{\ch}{\mathcal{H}}

\begin{document}


\title{Antichiral surface states and Su-Schrieffer-Heeger physics in rutile altermagnets}
\author{Sopheak Sorn}
\email{sopheak.sorn@kit.edu}
\affiliation{
Institute of Quantum Materials and Technology, Karlsruhe Institute of Technology, 
76131 Karlsruhe, Germany
}

\pacs{}
\date{\today}

\begin{abstract}
We study surface states and domain wall bound states in altermagnets using a rutile-lattice tight-binding model of electrons coupled to a N\'eel order. We discover that two symmetry-protected Weyl nodal lines in the bulk band structure can give rise to unconventional anti-chiral surface states---surface states from opposite surfaces propagate in a \emph{parallel} manner, as opposed to the anti-parallel manner for the more conventional chiral surface states. We also find that the anti-chiral surface states can be turned into chiral surface states upon changing the surface termination. The origin of the surface states, the dependence on the surface termination, and key features of domain wall bound states are explained using a map from the altermagnet to a family of a modified Su-Schrieffer-Heeger(SSH) chain and the associated bulk-boundary correspondence.
Our work reveals rutile altermagnets as a promising candidate among very few quantum materials that can support anti-chiral surface states. 
\end{abstract}

\maketitle

\begin{figure}[t]
    \centering
    \includegraphics[width=\linewidth]{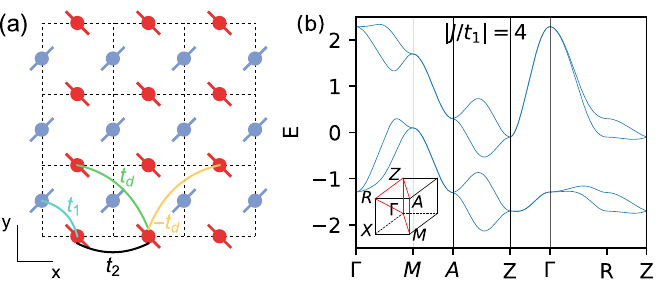}
    \caption{(a) Illustration of a rutile lattice featuring the two magnetic sublattices, A (blue dots) and B (red dots). They reside on different xy-planes and experience different local potentials due to non-magnetic $\mathcal{B}$ sites (not shown), whose impacts are effectively represented by the orientation of the tilted bars.
    (b) Tight-binding band structure for $\vec{N}_i = \hat z$, displaying Weyl nodal lines along $MA$ and $\Gamma Z$. ($t_1 = -0.2, t_2 = 0.05, t'_2 = -0.35, t_d = 0.15$ (comparable to the hopping parameters in the single-orbital model for RuO$_2$ from an \textit{ab initio} study \cite{Daniel2024}), $J = 0.8, \lambda = - 0.025$ and $\lambda' = 0.0125.$) }
    \label{fig:fig1}
\end{figure}

\noindent \textcolor{blue}{\textit{Introduction---}}
Recently, an emerging class of unconventional collinear antiferromagnets, known as altermagnet, has attracted a lot of attention due to their rich properties and their prospects in spintronic applications \cite{Jungwirth2022, Jungwirth2022B, Yao2024rev}. $\mathcal{AB}_2$ compounds featuring a three-dimensional(3D) rutile crystal structure have been shown to be a promising platform for altermagnetism. 
Examples of candidate materials include metallic RuO$_2$, insulating MnF$_2$, CoF$_2$, and ReO$_2$\cite{Yao2024rev}. 
Each unit cell contains two magnetic $\mathcal{A}$ elements residing on two sublattices whose local environments are made distinct from each other by the arrangement of the non-magnetic $\mathcal{B}$ elements.
The magnetic ordering is marked by the onset of a N\'eel order whose magnetic unit cell coincides with the unit cell of the lattice.
Remarkably, the two sublattices are not linked by an inversion operation, but rather linked by a four-fold screw rotation instead; see Fig.\ref{fig:fig1}(a). Consequently, the band structure in the altermagnetic state lacks the Kramers' degeneracy associated with $\mathcal{P}\mathcal{T}$ symmetry and exhibits a characteristic d-wave momentum-dependent spin splitting \cite{Hayami2019, Hayami2020, Jungwirth2022, Jungwirth2022B, Yao2024rev, Jan2019, Naka2019, Nakamura2016}.

Symmetry plays a crucial role in a diverse range of phenomena in rutile altermagnets. For instance, when the N\'eel vector points along the [001] crystallographic direction \cite{Weitering2017, Smejkal2020}, the corresponding nonsymmorphic magnetic space group P4$_{2'}$/mnm$'$ results in multiple Weyl nodal lines (WNL) in the band structure \cite{Yao2024, Fernandes2024pintchpoint, Fernandes2024mirror}. Moreover, the net magnetization and the linear-order anomalous Hall effect are constrained to be zero by the presence of two glide mirror symmetries \cite{Yao2024}. The symmetry admits a leading order parameter which displays the hallmark of a magnetic octupole moment \cite{Spaldin2024, Joerg2023, Rau2024, Fernandes2024pintchpoint}. Such an octupolar order parameter is closely connected to the rise of a nonlinear magneto-electric effect starting only at the second order \cite{Peters2024} and a nonlinear Hall response beginning only at the third order \cite{Ghorashi2024, Sorn2024}. It can also couple dynamically with the strain field, giving rise to a characteristic signature from a magnon-polaron hybridization \cite{Joerg2023}. The symmetry can be altered by tuning the N\'eel vector away from the [001] direction, which activates the linear-order anomalous Hall effect accompanied by a weak ferromagnetism or ferrimagnetism \cite{Rao1968, Comin2019, Smejkal2020B, Feng2022, Smejkal2022, Wang2023}.

The survey of the roles of symmetry brings about the following question: which physical phenomena emerge when the symmetry is reduced by inhomogeneities in the system? 
In this letter, we address this question focusing on the electronic spectral properties in the presence of inhomogeneity stemming from interfaces with vacuum and from magnetic domain walls(DW). To study this, we employ a rutile-lattice tight-binding model of electrons coupled to a N\'eel order on a slab geometry.

We uncover the existence of anti-chiral surface states(SS): SS on the opposite surfaces of the slab propagate in the \emph{same} direction. This is in contrast with the familiar case of chiral SS, such as the edge states in Chern insulators and the Fermi arc states in Weyl semimetals, which propagate in an antiparallel fashion between two opposite surfaces. Anti-chiral SS were first theoretically proposed in a variant of the Haldane honeycomb model where the Haldane flux pattern is modified\cite{Franz2018}.
However, so far, only few quantum materials have been suggested as their potential hosts, including twisted van der Waals multi-layers and Dirac systems where the anti-chiral SS are stabilized by the electron-phonon coupling\cite{Zilberberg2020, Foa2022, Luis2023, Li2024}.
Our work demonstrates that rutile altermagnets are an alternative quantum-material platform for their realization. In a large-coupling limit, we explain their origin from two WNL in the bulk using a map from the altermagnet to a family of modified SSH chains. The latter can carry a nontrivial topological invariant, resulting in edge states which are identifiable with the anti-chiral SS. Using the map, we also explain key features of DW bound states and how the anti-chiral SS can be switched into chiral SS upon altering the surface termination. We also show how the anti-chiral SS are stable against an applied Zeeman field and the tilting of the N\'eel vector into certain directions. Finally, we discuss the possibility of anti-chiral SS in nonrutile altermagnets.

\noindent \textcolor{blue}{\textit{Model---}} We consider a 3D electronic tight-binding model for rutile d-wave altermagnets introduced in Ref.\cite{Fernandes2024mirror, Daniel2024} as a minimal model to study topological properties in the 3D band structure. It is represented by the following Bloch Hamiltonian
\bea
\label{eq:model}
\ch &=& \ch_0 + \ch_{\rm soc},\\
\ch_0 &=& - 8t_1c_{x/2}c_{y/2}c_{z/2}\tau_x - 2t'_2 c_z\tau_0 - 2t_2 (c_x + c_y)\tau_0 \nonumber\\
&& - 4 t_d s_x s_y \tau_z + J \tau_z \vec{N}\cdot \vec{\sigma}, \\
\ch_{\rm soc} &=& - 8\lambda s_{z/2} (s_{x/2}c_{y/2} \sigma_x - s_{y/2}c_{x/2} \sigma_y)\tau_y\nonumber\\
&& + 16\lambda' c_{x/2}c_{y/2}c_{z/2}(c_x-c_y)\tau_y \sigma_z,
\eea
where $c_{\alpha/n}\equiv \cos(k_{\alpha}/n)$ and $s_{\alpha/n}\equiv \sin(k_{\alpha}/n)$. The basis vector is $(a_{\vec{k}\uparrow}, a_{\vec{k}\downarrow}, b_{\vec{k}\uparrow}, b_{\vec{k}\downarrow})^T$, where $a$ and $b$ are the annihilation operators on the sublattices A and B (see Fig.\ref{fig:fig1}(a).) The Pauli matrices, $\tau_i$ and $\sigma_i$, act on the sublattice and the spin index, respectively. The N\'eel vector for the i-th unit cell is defined by the difference of the magnetic dipole moments on the two sublattices: $\vec{N}_i = \vec{m}_{i,A} - \vec{m}_{i,B}$. $\ch_0$ encompasses spin-preserved hoppings and a coupling to a uniform N\'eel order $\vec{N}_i = \vec{N}$. $\ch_{\rm soc}$ includes hopping due to spin-orbit-coupling effects. Notably, a proper tight-binding description of a rutile altermagnet may require more structures in addition to Eq.\eqref{eq:model}. For instance, due to a multi-orbital character of RuO$_2$ near the Fermi level, four additional bands must be added to the model in Eq.\eqref{eq:model} \cite{Daniel2024}. However, our formulation can be accordingly generalized to the more complicated cases. 

Figure \ref{fig:fig1}(b) shows the band structure for $\vec{N} = \hat{z}$, corresponding to the magnetic point group $4'/mm'm$. It contains a two-fold rotation z-axis and mirror planes $\mathcal{M}_{x,y}$ associated with the nonsymmorphic glide mirrors, $\{\mathcal{M}_x|\frac{1}{2}\frac{1}{2}\frac{1}{2}\}$ and $\{\mathcal{M}_y|\frac{1}{2}\frac{1}{2}\frac{1}{2}\}$. There are Weyl nodes and WNL, whose symmetry protection has been discussed in Ref.\cite{Fernandes2024mirror, Fernandes2024pintchpoint}. In particular, the WNL along $\Gamma Z$ and $MA$ direction in Fig.\ref{fig:fig1}(b)  are protected by the two-fold z-axis rotation and the glide mirrors\cite{Fernandes2024mirror}. 
Next, we show that these WNL can give rise to the anti-chiral SS in a slab geometry. We emphasize that spin-orbit coupling (SOC) is important for the anti-chiral SS since the WNL are the results of SOC-induced gapping of predating nodal surfaces present in the absence of SOC \cite{Mazin2021}; for instance, when $\lambda = \lambda' = 0$, the relevant nodal surfaces are in the $k_zk_x$-plane and the $k_yk_z$-plane containing the $\Gamma$ point as well as the $k_zk_x$-plane and the $k_yk_z$-plane containing the M point.

\begin{figure}[t]
    \centering
    \includegraphics[width=\linewidth]{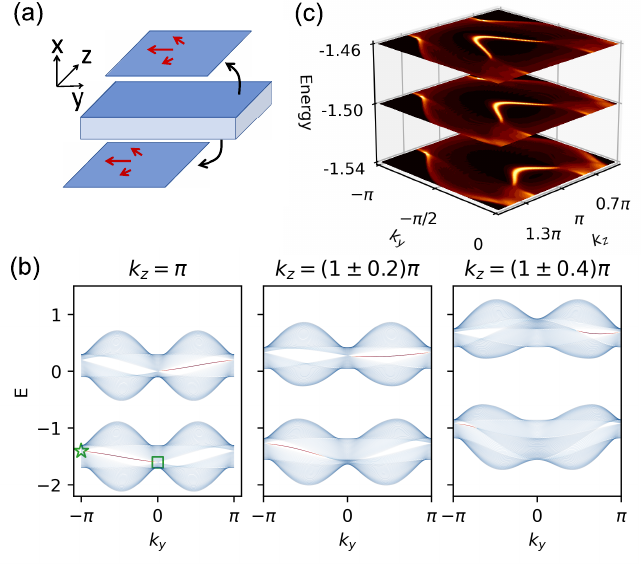}
    \caption{(a) Schematic illustration of the slab geometry, featuring anti-chiral SS (red arrows) which propagate predominantly in the negative y-direction. (b) Band structures of the slab geometry for $|J/t_1|=4$, highlighting the in-gap anti-chiral SS in red. The starting and ending points of the SS curves, e.g. the square and the star, descend from the WNL on $\Gamma Z$ and $MA$. (c) Contour plots showing that the anit-chiral SS, in bright yellow, form open arcs.}
    \label{fig:surface_state}
\end{figure}

\noindent \textcolor{blue}{\textit{Anti-chiral surface states---}} We consider a slab geometry with open boundaries along the x-direction as in Fig.\ref{fig:surface_state}(a). The crystal is terminated at A(B) sites for the bottom(top) surface, which is referred to as AB termination. Results for other types and other orientations of the surface termination can be generalized from this.
With the remaining two directions of translation invariance, the problem is mapped to one-dimensional(1D) chains parameterized by the crystal momenta $k_y$ and $k_z$ belonging to the surface Brillouin zone, $\,[-\pi, \pi) \times \,[-\pi, \pi)$.

\begin{figure}[t]
    \centering
    \includegraphics[width=0.95\linewidth]{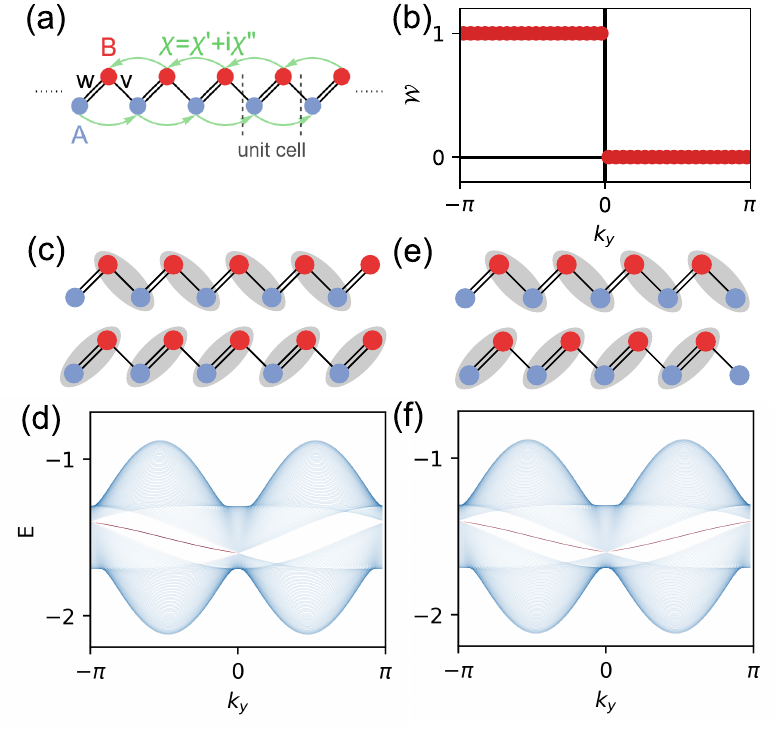}
    \caption{(a) A real-space representation of the modified SSH chain.
    (b) $k_y$ dependence of the $Z_2$ topological invariant $\mathcal{W}$ for the valence band of $\ch_{\rm SSH}$. (c) and (e) illustrate the dimerized limits for the AB and the AA surface termination, respectively. Unpaired outermost sites are identifiable with the SS. (d) and (f) are the band structures of the low-energy sector for the AB and the AA termination, respectively, for $k_z = \pi$. We observe a change from anti-chiral to chiral SS between the two types of surface termination.
    }
    \label{fig:ssh_physics}
\end{figure}

Figure \ref{fig:surface_state}(b) shows the band structure for a few values of $k_z$ for $\vec{N} = \hat{z}$. The blue continua correspond to the projection of the bulk bands onto the surface Brillouin zone. We observe energy gaps hosting in-gap SS which are exponentially localized to \emph{both} surfaces and represented by the doubly degenerate red curves. In particular, the curve connecting the star and the square in Fig.\ref{fig:surface_state}(b) consists of two SS---one for each surface. These SS propagate predominantly in the same negative y-direction, which can be seen from their slope. 
These are the anti-chiral SS, as introduced at the beginning.
Their unidirectional feature prevails for the whole portion of the parameter space $(k_y, k_z)$ where the SS are observed, as depicted in Fig.\ref{fig:surface_state}(c). There, the SS correspond to the bright doubly degenerate open arcs extending from $k_z < \pi$ to $k_z > \pi$. Indeed, they feature the same, predominantly negative group velocity $v_y$. The red arrows in Fig.\ref{fig:surface_state}(a) schematically illustrates the unidirectional behavior. Notably, the anti-chiral SS propagating in the positive y-direction also exist, but they reside at a higher energy, as seen in Fig.\ref{fig:surface_state}(b) for $0 < k_y < \pi$. The dependence of the anti-chiral SS on a few relevant model parameters can be found in the Appendix \ref{sec:appA}.

The anti-chiral SS create a nonzero contribution to the equilibrium charge current, which is necessarily canceled by another contribution from the bulk.
As mentioned before, such anti-chiral SS have been proposed in only a few solid-state systems \cite{Zilberberg2020, Foa2022, Li2024}. Here, we have clearly demonstrated rutile altermagnets as another natural host. Their physical origin turns out to arise from SSH physics and the associated bulk-boundary correspondence, which is discussed next.

\noindent \textcolor{blue}{\textit{Map to modified SSH chains---}}
The anti-chiral SS are intimately related to the WNL along $\Gamma Z$ and $MA$ of the bulk. For instance, the starting and the ending point of the surface-state curve in Fig.\ref{fig:surface_state}(b), i.e. the square and the star, descend from the nodal lines upon their projection onto the surface Brillouin zone. We will derive a map from our altermagnetic system to a family of modified SSH chains, which sheds light on the roles of the nodal lines and the rise of the anti-chiral SS.

To see the map, we proceed in two steps. Firstly, the map is most transparent in the large J-limit, where the full band structure decouples into two two-band sectors---one at a lower energy and one at a higher energy. The leading term in the Hamiltonian is the coupling to the N\'eel vector $\vec{N} = \hat{z}$, and the rest may be viewed as perturbations. From this, we derive effective Hamiltonians for the low-energy and the high-energy sector. Without loss of generality, we focus on the low-energy sector, which is spanned by the -1 eigenvectors of $\tau_z \sigma_z$, namely $\vec{v}_1 = (0, 1, 0, 0)^T$ and $\vec{v}_2 = (0, 0, 1, 0)^T$. In the first order of perturbation, the effective Hamiltonian is
\bea
\ch_L &=& -2t'_2 c_z - 2t_2 (c_x + c_y) - 4t_d s_x s_y \gamma_z - J \nonumber \\
&&- 8 \lambda s_{z/2} (s_{x/2}c_{y/2} \gamma_y - s_{y/2}c_{x/2} \gamma_x), 
\eea
where the Pauli matrices $\gamma_i$ act on the basis $\{\vec{v}_l\}$.

In the final step, we view $k_y$ and $k_z$ in $\ch_L$ as tuning parameters which can be absorbed into the original tight-binding parameters. 
The outcome is a modified SSH chain (see Fig.\ref{fig:ssh_physics}(a)):
\bea
H_{\rm SSH} &=& \sum_x \left(w c^{\dagger}_{A, x} c_{B, x} + v c^{\dagger}_{B, x}c_{A, x+1}\right) + \nonumber \\
&& \sum_x \left[ \chi c^{\dagger}_{A, x} c_{A, x+1} + \chi^* c^{\dagger}_{B, x} c_{B, x+1} \right] + \text{h.c.}
\eea
$ w = 4\lambda s_{z/2}(s_{y/2} + c_{y/2}), v = 4\lambda s_{z/2}(s_{y/2} - c_{y/2}), \text{ and } \chi = -t_2 + i 2t_d s_y$.
We have ignored terms in $\ch_L$ that do not depend on $k_x$. This is benign for the SS discussion thanks to their topological origin. We have also performed a gauge transformation, $\mathcal{U}^{\dagger}\ch_L \mathcal{U}$, where $\mathcal{U} = e^{-ik_x/4} \left(\cos \frac{k_x}{4} \gamma_0 + i \sin \frac{k_x}{4} \gamma_z\right)$. 
In reciprocal space,
\bea
\ch_{\rm SSH} &=& \begin{pmatrix}
\chi e^{ik_x} + \chi^* e^{-ik_x} & w + v e^{ik_x} \\
w + v e^{-ik_x} & \chi e^{-ik_x} + \chi^* e^{ik_x}
\end{pmatrix}.
\label{eq:ssh_recip}
\eea

The nonzero $\chi$ differentiates our model from the original SSH chain \cite{SSH}. The SSH chain is known to be a bulk obstructed atomic insulator, which is topologically trivial in the sense that it is smoothly connected to an atomic limit \cite{Cano2018, Vishwanath2018, Queiroz2021, Bernevig2024, Slager2017}. Nevertheless, it supports nontrivial SS arising from a nontrivial bulk topological invariant, in accordance with a bulk-boundary correspondence \cite{Bardarson2017}. 
Defining such a topological invariant generally requires some cares, as addressed in Ref.\cite{Bardarson2017}.
The relevant topological invariant for our consideration is the following bulk Z$_2$ index: 
\bea
\mathcal{W} &=& \gamma /\pi \text{ mod } 2.
\eea
$\gamma = i \oint d k_x \bra{\alpha(k_x; k_y, k_z)} \partial_{k_x} \ket{\alpha(k_x; k_y, k_z)}$, $\ket{\alpha}$ is the eigenstate for the valence band of $\ch_{\rm SSH}$ Eq. \eqref{eq:ssh_recip}, assumed to be smooth and periodic, $\ket{\alpha(k_x)} = \ket{\alpha(k_x + 2\pi)}$. $\gamma$ is the so-called intercellular part of the original Zak phase that enters the bulk-boundary correspondence \cite{Bardarson2017, Slager2019}. 

Figure \ref{fig:ssh_physics}(b) shows the $k_y$ dependence of $\mathcal{W}$, which jumps at $k_y = 0$ and $\pi$ due to the bulk gap closing associated with the $\Gamma Z$ and $MA$ WNL(see Fig. \ref{fig:fig1}(b)). For $-\pi < k_y < 0$, the nontrivial $\mathcal{W}$ and the bulk-boundary correspondence indeed explain the presence of SS in Fig.\ref{fig:ssh_physics}(d). This provides a clear understanding of the origin of the anti-chiral SS from the bulk topological invariant.
Next, we discuss their dependence on how the surfaces are terminated.

\noindent \textcolor{blue}{\textit{Dependence on surface termination---}} There are four types of [100] surface termination: AB, AA, BA, and BB. Remarkably, we find that the anit-chiral SS are supported only in the AB and BA terminations, while they become chiral for the AA and BB terminations; see Fig. \ref{fig:ssh_physics} (d) and (f). To understand this feature, we analyze the electronic behaviors in the representative cases of AB and AA terminations. Generalization to the other surface terminations is possible.

The discussion can be made transparent by considering two dimerized limits of Eq.\eqref{eq:ssh_recip}: (i) $|v| \gg |w|, |\chi|$ and (ii) $|w| \gg |v|,|\chi|$. Although the dimerized limits may not be reached in general by the altermagnet, they define equivalence classes and serve as abstract reference points which are smoothly deformed into the effective SSH chains in Eq.\eqref{eq:ssh_recip} without closing the energy gap between the two bands. We then use the adiabaticity argument to elucidate the existence of the SS of the altermagnet from the SS in the associated dimerized limits.

The large $|v|$ in (i) leads to a dimerization between the A and B sublattices of neighboring unit cells, defined in Fig.\ref{fig:ssh_physics}(a). Such a process is illustrated in the top row of Fig.\ref{fig:ssh_physics}(c) and is smoothly connected with our case with $\mathcal{W} = 1$. The leftmost A site and the rightmost B site of the AB open chain are left unpaired and hence identifiable with the SS. In contrast, our case with $\mathcal{W} = 0$ is smoothly connected with the dimerized limit (ii), where $|w|$ dominates so that A and B sublattices of the same unit cells form dimers instead. This is illustrated in the bottom row of Fig.\ref{fig:ssh_physics}(c), where the outermost sites participate in dimerization. Therefore, the SS are expected to be absent. The dimerization picture correctly explains the anti-chiral SS in the AB slab in Fig.\ref{fig:ssh_physics}(d) for $k_z = \pi$, where the window of $-\pi<k_y<0$ and $0<k_y<\pi$ correspond to the dimerized limit (i) and (ii), respectively.

We now deploy the dimerization picture for the AA slab, where the outermost sites are A sites; see Fig.\ref{fig:ssh_physics}(e). By viewing the open chains in panel (e) as resulted from removing the rightmost B sites in panel (c), we see that the top(bottom) row in Fig.\ref{fig:ssh_physics}(e) corresponds to the dimerized limit (i) ((ii)). The SS are derived from the leftmost(rightmost) unpaired A site. This implies that, for the altermagnet and for $k_z = \pi$, the SS are present in both windows: $-\pi < k_y < 0$ and $0 < k_y < \pi$. This is in an excellent agreement with the numerical results in Fig.\ref{fig:ssh_physics}(f). We note that the overall sign of the group velocity of the SS is mainly determined by the energy offset between the gapless points at $k_y = 0$ and $\pi$.

The SS of the AA slab are chiral.
By altering one outermost layer of the AB slab to obtain the AA slab, we switch from anti-chiral to chiral SS. Moreover, the contribution from the SS to the equilibrium current has reduced, which subsequently alters the spatial distribution of the equilibrium current in the bulk. 
The change of the equilibrium current distribution in this context is left for future investigations. We now turn bound states at altermagnetic DWs. 

\begin{figure}[t]
    \centering
    \includegraphics[width=\linewidth]{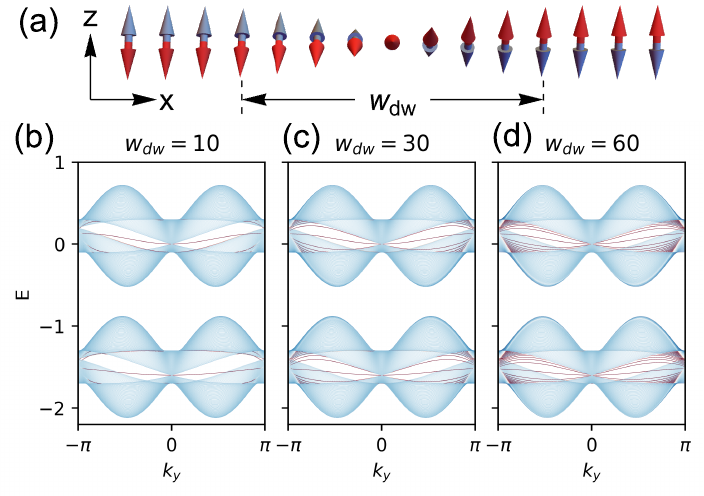}
    \caption{(a) Illustration of a DW of the N\'eel vector. Each blue(red) arrow denotes the dipole moment of the A(B) sublattice of a unit cell. (b) Dependence of the band structure at $k_z = \pi$ on the width of the DW, $w_{\rm dw}$. The DW bound states correspond to the in-gap states (red curves), whose number increases with $w_{\rm dw}$. When the DW states pass through the continuum bulk states, they hybridize with one another. 
    }
    \label{fig:dw}
\end{figure}

\noindent \textcolor{blue}{\textit{Bound states at altermagnetic domain walls---}} We consider the AB slab geometry with a Bloch-typed DW of the N\'eel vector, wherein the N\'eel vector of the i-th unit cell has the form: $\vec{N}_i = (0, \sin \theta_i, \cos \theta_i)^T$. For the left domain $i \leq -w_{\rm dw}/2$, $\theta_i = 0$, and for the right domain $i \geq w_{\rm dw}/2$, $\theta_i = \pi$.
In the DW region $-w_{\rm dw}/2<i<w_{\rm dw}/2$, $\theta_i = \pi(2i+w_{\rm dw})/2w_{\rm dw}$.
$w_{\rm dw}$ denotes the width of the DW depicted in Fig.\ref{fig:dw}(a). 

Figure \ref{fig:dw}(b)-(d) show the band structures for a few values of $w_{\rm dw}$, featuring (i) continua of bulk states in blue, (ii) in-gap DW bound states in red, and (iii) in-gap SS (omitted.) The sharp DW produces mainly one DW bound state within each energy gap, as seen in Fig.\ref{fig:dw}(b). They have a topological origin, for they can be regarded as the anti-chiral SS of the left and the right magnetic domains. Upon increasing $w_{\rm dw}$, more bound states appear, as seen in panel (c) and (d), which is independent of the topological property of the system. 
Such behaviors can be understood using a continuum approximation which maps the low-energy sector of the DW problem in the large $|J|$-limit (same for the high-energy sector) into a 1D Dirac fermion model featuring a nonuniform mass (see Appendix \ref{sec:appB} for a derivation):
\bea
H_{\rm dw} &=& v_F\left[i\partial_x \tilde{\gamma}_y + k_y \cos \theta(x)\tilde{\gamma}_x\right],
\eea
where $v_F = 4\lambda \sin \frac{k_z}{2}$ and $\tilde{\gamma}_i's$ are Pauli matrices. The mass $v_F k_y\cos\theta(x)$ depends on $\theta(x)$, a continuum version of $\theta_i$. The eigenvalues of $H_{\rm dw}$ describe the energy spectrum close to $k_y = 0$ at a fixed $k_z$. To gain further insight, we consider a specific DW profile, $\cos \theta(x) = \tanh \beta x$, and employ Bohr-Sommerfeld quantization condition to obtain the number of DW bound states for the limit $\beta/k_y \rightarrow 0$(see Appendix \ref{sec:appB} for details):
\bea
N_b &\approx& 2\sqrt{\frac{k_y^2 + |\beta k_y|}{\beta^2}}. 
\label{eq:bound}
\eea
Decreasing $\beta$ in Eq.\eqref{eq:bound} amounts to increasing $w_{\rm dw}$, which correctly explains the increasing number of DW bound states while going from panel (b) to (d) in Fig.\ref{fig:dw}. For a fixed $\beta$, the growing $N_b$ with larger $|k_y|$ explains the trend most obvious in Fig.\ref{fig:dw}(d), where more DW bound states emerge out of the bulk-state continua as one moves away from $k_y = 0$.

\noindent\textcolor{blue}{\textit{Discussion---}}
Firstly, the anti-chiral SS and their robustness are intertwined with the presence of the symmetry-protected WNL which are stable against an external Zeeman field in certain high-symmetry directions and the tilting of $\vec N$ away from z-axis within certain high-symmetry planes; see Appendix \ref{sec:appC} and also Ref.\cite{Fernandes2024pintchpoint}. In particular, an external Zeeman field in the x-axis or the y-axis does not gap out the WNL but rather causes them to move off the high-symmetry paths in the reciprocal space. In such cases, the anti-chiral SS persist. Otherwise, the WNL are gapped out, and the existence of the anti-chiral SS is no longer guaranteed.

Secondly, as mentioned before, SOC is essential since it gaps out nodal surfaces to produce the WNL. In addition, a larger SOC leads to a more pronounced gap within which the anti-chiral SS reside. As a result, the anti-chiral SS become more spatially localized and can be better resolved from the bulk states in the reciprocal space; see Fig.\ref{fig:lambda} in Appendix \ref{sec:appA}.

Lastly, a group-theory-based study shows that similar WNL exist in a wide range of altermagnets, which includes the rutiles \cite{Fernandes2024pintchpoint}. With such an analogous starting point, anti-chiral SS could be realized in other altermagnets if situations similar to the rutiles also arise: WNL arrange themselves into pairs. WNL in each pair are parallel to each other so that a slice of the 3D Hamiltonian $\ch(\vec k)$ in the plane perpendicular to the nodal lines can be viewed as an effective 2D Hamiltonian hosting a pair (or pairs) of Dirac cones. The Dirac cones may lead to anti-chiral SS if there is an energy offset between them. This is the case since each Dirac point corresponds to a topological transition within the SSH framework, as in Fig.\ref{fig:ssh_physics}(b). Then our results and insights from the rutiles could also be applicable. We leave this prospect for future investigation.

\noindent \textcolor{blue}{\textit{Summary---}} Our work demonstrates that rutile altermagnets support exotic anti-chiral SS, which are rare and desirable in quantum materials. Their origin comes from WNL in the bulk bands and is explained using the SSH mapping as well as its bulk-boundary correspondence. In a broader context, our results contribute to the fundamental understanding of altermagnetic systems by establishing a connection between different realms of condensed matter physics. 
Our work motivates further theoretical and experimental investigation to look for similar anti-chiral SS in other altermagnetic systems, to examine their robustness against disorder, and to confirm our proposal experimentally, e.g. via angular-resolved photo-emission spectroscopy or quasi-particle inference patterns from scanning tunneling microscopy.
It also suggests an exciting outlook to explore a potential connection between rutile altermagnets and graphene since the latter is similarly linked to the SSH physics \cite{Montambaux2011}.

\noindent \textcolor{blue}{\textit{Acknowledgment---}} The author would like to thank Markus Garst, Iksu Jang, Volodymyr Kravchuk and Geremia Massarelli for very helpful discussions and comments. The author also thanks Christian Pfleiderer, Wolfgang Simeth, and Andreas Bauer for many stimulating discussions in related projects. The author is supported by the Deutsche Forschungsgemeinschaft through TRR 288
Grant No. 422213477 (Project No. A11).

\appendix

\section{Dependence of anti-chiral surface states on model parameters}
\label{sec:appA}
\begin{figure}[t]
	\centering
	\includegraphics[width=0.98\linewidth]{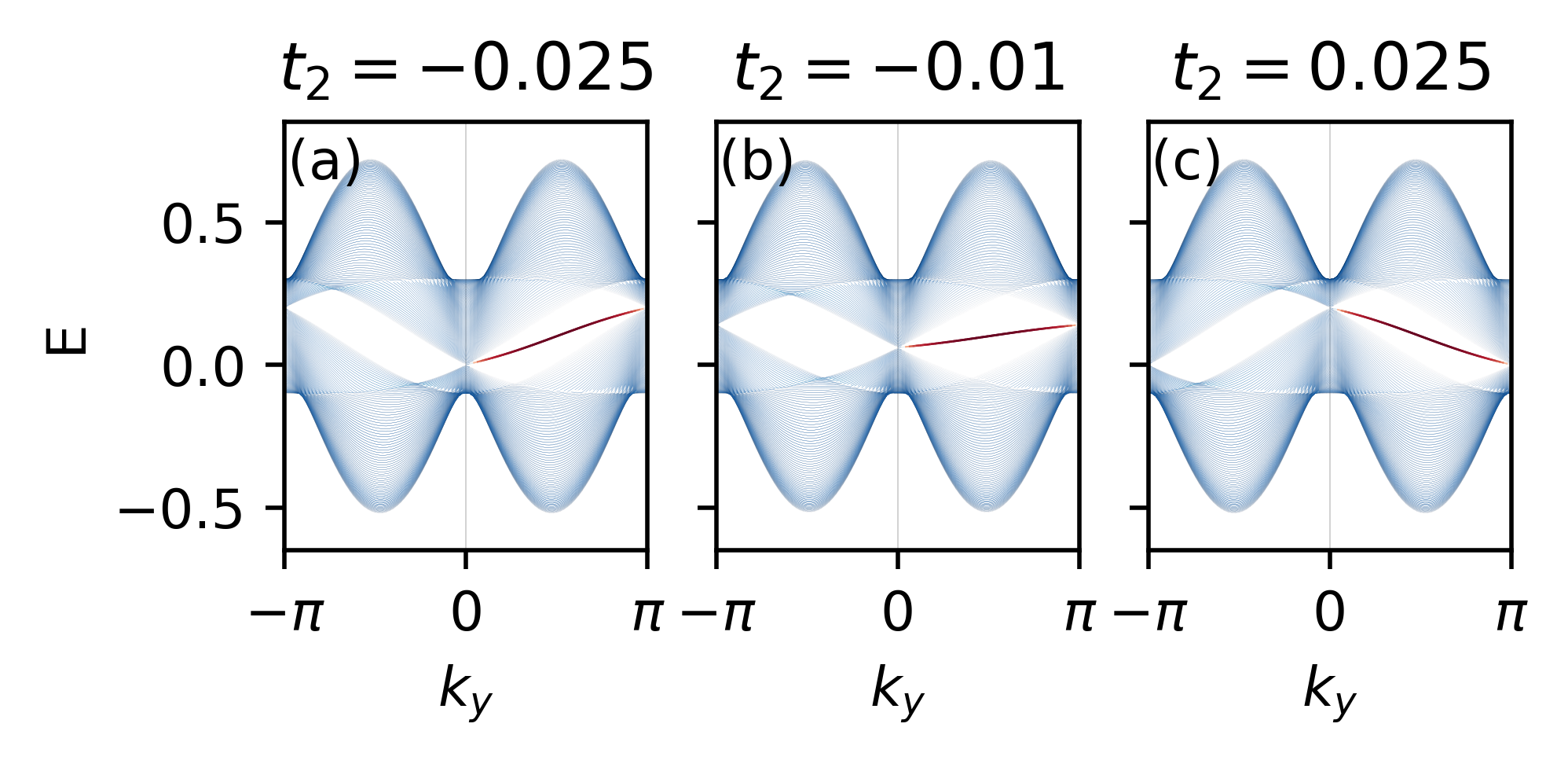}
	\caption{By changing $t_2$, one can tune the slope of the anti-chiral SS. Here, we choose $k_z = \pi$. The rest of the model parameters is kept at the values given in Fig.1 of the main text.} 
	\label{fig:t2}
\end{figure}

Figure \ref{fig:t2} shows how $t_2$ tunes the slope of the anti-chiral surface states(SS). When $t_2 = 0$, the slope is more or less zero, marking the transition from a positive to a negative slope. We note that $t_2$ only influences the energy eigenvalues of $\ch(\vec k)$ but not the eigenvectors 
since it appears in front of the identity matrix, $\tau_0 \sigma_0$, in $\ch$.

\begin{figure}[t]
	\centering
	\includegraphics[width=0.98\linewidth]{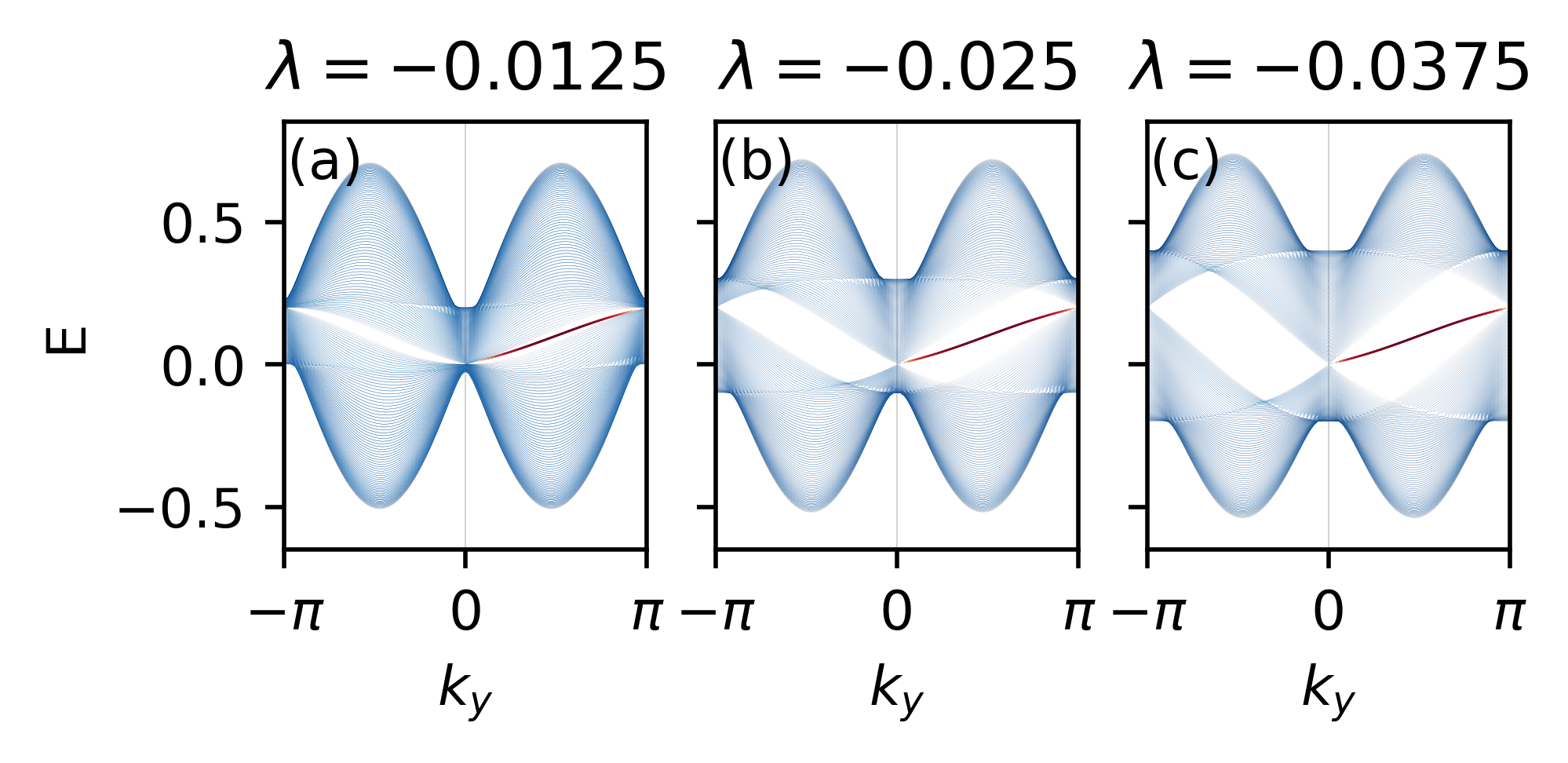}
	\caption{By changing $\lambda$, one can tune the size of the gap wherein the anti-chiral SS reside. Here, we show the cases with $k_z = \pi$. The rest of the model parameters is kept at the values given in Fig.1 of the main text.}  
	\label{fig:lambda}
\end{figure}

Figure \ref{fig:lambda} shows how the energy gap, within which the anti-chiral SS reside, depends on the strength of the spin-orbit-coupling parameter $\lambda$. As usual, a more pronounced energy gap due to a large $\lambda$ results in a more spatially localized profile of the anti-chiral SS and a better resolution between the anti-chiral SS and the bulk states in the reciprocal space.

\section{Properties of bound states at altermagnetic domain walls}
\label{sec:appB}
The domain wall (DW) problem of the altermagnet can be mapped to a DW problem of the generalized SSH chain in the large J-limit, wherein the hopping integrals become spatially dependent. We find that its continuum version can provide a qualitative understanding of our numerical results in Fig.4(b)-(d) in the main text, so this will be the focus here. We analyze the continuum version by performing the following steps: (i) obtain a long-wavelength limit of Eq.(1) of the main text by expanding around $(k_x, k_y) = (0, 0)$ in the linear order of $k_x$ and $k_y$, (ii) exchange $k_x$ for $-i\partial_x$, (iii) perform a local spin rotation represented by a spatially dependent unitary operation, $\mathcal{U}_{\mathcal{R}}(x) = \vec{\sigma}\cdot (0, \sin \frac{\theta}{2}, \cos \frac{\theta}{2})$, where $\theta(x)$ is a continuous representation of $\theta_i$. This simplifies the J-coupling term $J \tau_z \vec{N}(x) \cdot \vec{\sigma} \mapsto J \tau_z \sigma_z$, at the expense of more complicated kinetic terms. Finally, (iv) we take the large-J limit, akin to the previous sections. We obtain an approximated Hamiltonian for the low-energy sector:
\bea
H_{\rm dw} &\sim & v_F \left[ i \partial_x \gamma_y + k_y \cos \theta(x) \gamma_x \right],
\eea
where $v_F = 4\lambda \sin \frac{k_z}{2}$. The eigenvalue problem for a general $\theta(x)$ is given by 
\bea
H_{\rm dw} \vec{\psi} = E \vec{\psi},
\eea
where $\vec{\psi} = (\psi_1, \psi_2)^T$. Indeed, our DW problem can be viewed as a 1D Dirac fermion with a spatially varying mass term given by $k_y \cos\theta(x)$. For a step-like DW,
\bea
\cos\theta(x) &=& 2 \Theta(x) - 1,
\eea
where $\Theta$ is the heavy-side step function, $\vec{\psi}$ is the Jackiw-Rebbi bound state solution \cite{Shen2012}. The bound state is related to the topological properties of the left and the right magnetic domains, as discussed in the main text. Next, we move on to estimate the number of bound states.

For a generic DW profile, the eigenvalue problem can be solved by solving the following differential equation for each component,
\bea
v_F^2\left[-\partial^2_x + k_y^2 \cos^2\theta + S_{\lambda} k_y \partial_x \cos\theta \right] \psi_{\lambda} &=& E^2 \psi_{\lambda},
\label{eq:well}
\eea
where $S_{\lambda} = \pm 1$ for $\lambda = 1, 2$, respectively.
This amounts to a problem for a free particle in a potential well $V_{\lambda}(x) = k_y^2\cos^2\theta + S_{\lambda} k_y \partial_x \cos\theta$.
For a reasonable, smooth DW profile, the potential always has a well-like dip and admits at least one bound state either for $\lambda = 1 \text{ or } 2$, depending on the sign of $S_{\lambda} k_y \partial_x \cos \theta$. 

\begin{figure}[t]
    \centering
    \includegraphics[width=0.7\linewidth]{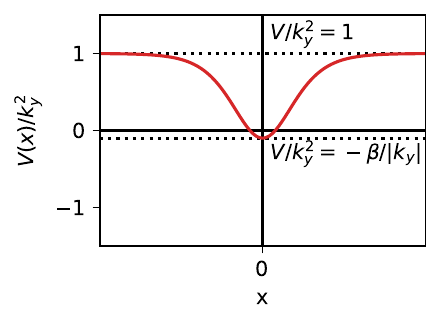}
    \caption{Illustration of the potential well $V(x)$ which traps bound states.}
    \label{fig:pot_well}
\end{figure}

Consider the following functional form $\cos \theta = \tanh \beta x$ for the DW profile, where $\beta >0 $ tunes the width of the DW region. Figure \ref{fig:pot_well} shows the profile of the potential well, featuring a height set by $k_y^2$ and a dip at $-|k_y \beta|$ (we focus on the case $S_{\lambda}k_y < 0$, which supports the bound states.) In the limit of $\beta/|k_y| \rightarrow 0$, the bottom of the potential well is more or less at zero energy. In this limit, we can relate the number of bound states $n_b$ of Eq.\eqref{eq:well} to the number of bound states $N_b$ for the altermagnetic DW at a given $k_y$:
\bea
N_b &\approx& 2 n_b.
\eea
The factor 2 accounts for the positive and negative value of $E$ for a given $\varepsilon$.
To obtain $n_b$, we employ the Bohr-Sommerfeld quantization condition. $n_b$ is the number of quantized energy levels for $\varepsilon < k_y^2$ for the Hamiltonian in Eq.\eqref{eq:well}.
The Bohr-Sommerfeld quantization condition implies that
\bea
n_b h &=& 2 \int_{-\infty}^{\infty} p(\varepsilon = k_y^2) dx, 
\eea
where $\pm \infty$ are the classical turning points for the energy $\varepsilon = k_y^2$, which is the height of the potential well. The momentum $p = \hbar\sqrt{\varepsilon - V(x)}$. Plugging the form of $V(x)$, we obtain 
\bea
n_b &=& \sqrt{\frac{k_y^2 + |\beta k_y|}{\beta^2}}.
\eea
From this, we arrive at the number of bound states $N_b$
\bea
N_b &\approx & 2 \sqrt{\frac{k_y^2 + |\beta k_y|}{\beta^2}}.
\eea
It entails more bound states either by decreasing $\beta$ or by increasing $|k_y|$. Such a behavior is in a good agreement with the numerical results in Fig.4(b)-(d) in the main text. Lowering $\beta$ amounting to increasing $w_{\rm dw}$ indeed increases the number of bound states. Meanwhile, more bound states appear as we move away from $k_y = 0$, which is most apparent in panel (d), where more bound states emerge out of the blue continua away from $k_y = 0$.

\begin{figure}[t]
	\centering
	\includegraphics[width=0.98\linewidth]{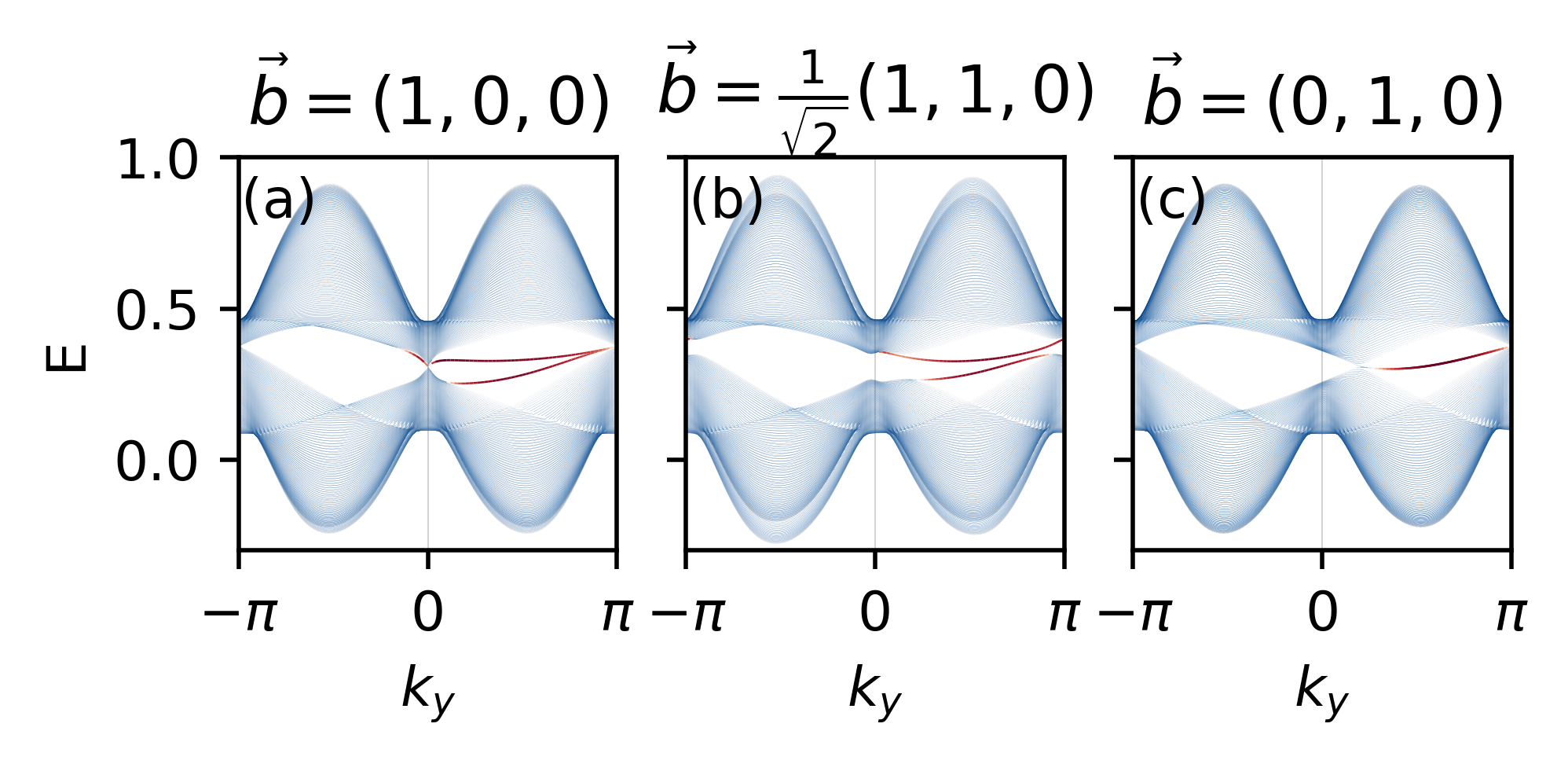}
	\caption{Impacts of the external Zeeman field $\vec b$ on the nodal points at $k_y = 0, \pi$ (which descend from the WNL along $\Gamma Z$ and $MA$ lines when $\vec b = 0$) and the anti-chiral SS (shown in red). The direction of the $\vec b$ is given at the top of each panel. In (a) and (c), applying $\vec b$ along the x-axis or the y-axis does not gap out the nodal points, and the anti-chiral SS persist. At most, $\vec b$ changes either the slopes of the anti-chiral SS or the position of the nodal points; for instance, see (c) where the position of the nodal point at $k_y = 0$ has been shifted. In (b), applying $\vec b = \frac{1}{\sqrt{2}} (1, 1, 0)^T$ gaps out the nodal points. The existence of the anti-chiral SS is not longer guaranteed by any nodal structures. This figure is obtained from adding $\ch_{\rm ext}^{(1)} + \ch_{\rm ext}^{(2)}$ to the results of Figure 2(b) of the main text, and we have chosen here $k_z = 1.2\pi$, $\alpha_1 = 0.1$ and $\alpha_2 = 0.25$.} 
	\label{fig:robustness}
\end{figure}

\section{Stability of Weyl nodal lines and anti-chiral surface states}
\label{sec:appC}
In this section, we consider how the anti-chiral SS behave under the impact of a Zeeman field and the tilting of $\vec N$ away from the z-axis. These two perturbations produce a similar effect, which can be seen from the Landau theory for rutile altermagnets. The theory permits the following symmetry-allowed Lifshitz invariant \cite{Rau2024, Fernandes2024mirror}
\bea
L &\sim & N_x b_y + N_y b_x,
\label{eq:Lifshitz}
\eea
where $\vec b = (b_x, b_y, b_z)^T$ represents the Zeeman field, and $\vec N = (N_x, N_y, N_z)^T$ denotes the N\'eel vector. Equation \eqref{eq:Lifshitz} implies that applying $\vec b$ can induce a tilting of $\vec N$, or alternatively, tilting $\vec N$ away from the z-axis induces an internal $\vec b$ field which manifests as a net magnetization proportional to the induced $\vec b$---a weak ferromagnetism phenomenon. Since the intertwined effect of $\vec b$ and tilting $\vec N$ takes place spontaneously, we consider them on an equal footing by considering a combined perturbation $\ch_{\rm ext}^{(1)} + \ch_{\rm ext}^{(2)}$ to our model Hamiltonian $\ch$ in the main text
\bea
\label{eq:ext1}
\ch^{(1)}_{\rm ext} & = & \alpha_1 \vec{\sigma}\cdot \vec b,\\
\label{eq:ext2}
\ch_{\rm ext}^{(2)} &=& \alpha_2 \tau_z (b_x \sigma_y + b_y \sigma_x),
\eea
where $\alpha_{1,2}$ are coupling coefficients which are not fixed by the symmetry argument. $\ch_{\rm ext}^{(1)}$ describes a Zeeman coupling to the conduction electrons. $\ch_{\rm ext}^{(2)}$ describes how the induced tilting of $\vec N$ is coupled with the conduction electrons. Note that we have replaced $N_x$ and $N_y$ in $\ch_{\rm ext}^{(2)}$ by $b_y$ and $b_x$, respectively. Alternatively, if we consider the tilting of $\vec N$ as the starting point and $\vec b$ as an induced field, we can replace $b_x$ and $b_y$ in Eq.\eqref{eq:ext1} and \eqref{eq:ext2} by $N_y$ and $N_x$, respectively.

Figure \ref{fig:robustness} (a) and (c) show how the anti-chiral SS and the nodal points near $k_y = 0, \pi$ associated with WNL persist in the presence of a Zeeman field $\vec b = (1, 0, 0)^T$ and $\vec b = (0, 1, 0)^T$ or when $\vec N$ tilts in the zx-plane and the yz-plane. In these cases, only the slope of the anti-chiral SS and the position of the nodal point near $k_y = 0$ are modified. However, applying $\vec b$ or tilting $\vec N$ in any other directions will gap out the nodal points, as seen in Fig.\ref{fig:robustness}(b), and the existence of the anti-chiral SS is not guaranteed by any nodal structures.
Our numerical results agree well with the general behaviors of the WNL in a variety of altermagnets in response to an external Zeeman field, which is studied in great detail in Ref. \cite{Fernandes2024pintchpoint}; see specifically the discussion in their Section IV.

\bibliographystyle{unsrtnat}
\bibliography{refs}

\begin{thebibliography}{42}
\providecommand{\natexlab}[1]{#1}
\providecommand{\url}[1]{\texttt{#1}}
\expandafter\ifx\csname urlstyle\endcsname\relax
  \providecommand{\doi}[1]{doi: #1}\else
  \providecommand{\doi}{doi: \begingroup \urlstyle{rm}\Url}\fi

\bibitem[\ifmmode~\check{S}\else \v{S}\fi{}mejkal
  et~al.(2022{\natexlab{a}})\ifmmode~\check{S}\else \v{S}\fi{}mejkal, Sinova,
  and Jungwirth]{Jungwirth2022}
Libor \ifmmode~\check{S}\else \v{S}\fi{}mejkal, Jairo Sinova, and Tomas
  Jungwirth.
\newblock Beyond conventional ferromagnetism and antiferromagnetism: A phase
  with nonrelativistic spin and crystal rotation symmetry.
\newblock \emph{Phys. Rev. X}, 12:\penalty0 031042, Sep 2022{\natexlab{a}}.
\newblock \doi{10.1103/PhysRevX.12.031042}.
\newblock URL \url{https://link.aps.org/doi/10.1103/PhysRevX.12.031042}.

\bibitem[\ifmmode~\check{S}\else \v{S}\fi{}mejkal
  et~al.(2022{\natexlab{b}})\ifmmode~\check{S}\else \v{S}\fi{}mejkal, Sinova,
  and Jungwirth]{Jungwirth2022B}
Libor \ifmmode~\check{S}\else \v{S}\fi{}mejkal, Jairo Sinova, and Tomas
  Jungwirth.
\newblock Emerging research landscape of altermagnetism.
\newblock \emph{Phys. Rev. X}, 12:\penalty0 040501, Dec 2022{\natexlab{b}}.
\newblock \doi{10.1103/PhysRevX.12.040501}.
\newblock URL \url{https://link.aps.org/doi/10.1103/PhysRevX.12.040501}.

\bibitem[Bai et~al.(2024)Bai, Feng, Liu, Šmejkal, Mokrousov, and
  Yao]{Yao2024rev}
Ling Bai, Wanxiang Feng, Siyuan Liu, Libor Šmejkal, Yuriy Mokrousov, and Yugui
  Yao.
\newblock Altermagnetism: Exploring new frontiers in magnetism and spintronics.
\newblock \emph{Advanced Functional Materials}, n/a\penalty0 (n/a):\penalty0
  2409327, 2024.
\newblock \doi{https://doi.org/10.1002/adfm.202409327}.
\newblock URL
  \url{https://onlinelibrary.wiley.com/doi/abs/10.1002/adfm.202409327}.

\bibitem[Hayami et~al.(2019)Hayami, Yanagi, and Kusunose]{Hayami2019}
Satoru Hayami, Yuki Yanagi, and Hiroaki Kusunose.
\newblock Momentum-dependent spin splitting by collinear antiferromagnetic
  ordering.
\newblock \emph{Journal of the Physical Society of Japan}, 88\penalty0
  (12):\penalty0 123702, 2019.
\newblock \doi{10.7566/JPSJ.88.123702}.
\newblock URL \url{https://doi.org/10.7566/JPSJ.88.123702}.

\bibitem[Hayami et~al.(2020)Hayami, Yanagi, and Kusunose]{Hayami2020}
Satoru Hayami, Yuki Yanagi, and Hiroaki Kusunose.
\newblock Bottom-up design of spin-split and reshaped electronic band
  structures in antiferromagnets without spin-orbit coupling: Procedure on the
  basis of augmented multipoles.
\newblock \emph{Phys. Rev. B}, 102:\penalty0 144441, Oct 2020.
\newblock \doi{10.1103/PhysRevB.102.144441}.
\newblock URL \url{https://link.aps.org/doi/10.1103/PhysRevB.102.144441}.

\bibitem[Ahn et~al.(2019)Ahn, Hariki, Lee, and Kune\ifmmode~\check{s}\else
  \v{s}\fi{}]{Jan2019}
Kyo-Hoon Ahn, Atsushi Hariki, Kwan-Woo Lee, and Jan Kune\ifmmode~\check{s}\else
  \v{s}\fi{}.
\newblock Antiferromagnetism in ${\mathrm{ruo}}_{2}$ as $d$-wave pomeranchuk
  instability.
\newblock \emph{Phys. Rev. B}, 99:\penalty0 184432, May 2019.
\newblock \doi{10.1103/PhysRevB.99.184432}.
\newblock URL \url{https://link.aps.org/doi/10.1103/PhysRevB.99.184432}.

\bibitem[Naka et~al.(2019)Naka, Hayami, Kusunose, Yanagi, Motome, and
  Seo]{Naka2019}
Makoto Naka, Satoru Hayami, Hiroaki Kusunose, Yuki Yanagi, Yukitoshi Motome,
  and Hitoshi Seo.
\newblock Spin current generation in organic antiferromagnets.
\newblock \emph{Nature Communications}, 10\penalty0 (1):\penalty0 4305, Sep
  2019.
\newblock ISSN 2041-1723.
\newblock \doi{10.1038/s41467-019-12229-y}.
\newblock URL \url{https://doi.org/10.1038/s41467-019-12229-y}.

\bibitem[Noda et~al.(2016)Noda, Ohno, and Nakamura]{Nakamura2016}
Yusuke Noda, Kaoru Ohno, and Shinichiro Nakamura.
\newblock Momentum-dependent band spin splitting in semiconducting mno2: a
  density functional calculation.
\newblock \emph{Phys. Chem. Chem. Phys.}, 18:\penalty0 13294--13303, 2016.
\newblock \doi{10.1039/C5CP07806G}.
\newblock URL \url{http://dx.doi.org/10.1039/C5CP07806G}.

\bibitem[Roig et~al.(2024)Roig, Kreisel, Yu, Andersen, and
  Agterberg]{Daniel2024}
Merc\`e Roig, Andreas Kreisel, Yue Yu, Brian~M. Andersen, and Daniel~F.
  Agterberg.
\newblock Minimal models for altermagnetism.
\newblock \emph{Phys. Rev. B}, 110:\penalty0 144412, Oct 2024.
\newblock \doi{10.1103/PhysRevB.110.144412}.
\newblock URL \url{https://link.aps.org/doi/10.1103/PhysRevB.110.144412}.

\bibitem[Berlijn et~al.(2017)Berlijn, Snijders, Delaire, Zhou, Maier, Cao, Chi,
  Matsuda, Wang, Koehler, Kent, and Weitering]{Weitering2017}
T.~Berlijn, P.~C. Snijders, O.~Delaire, H.-D. Zhou, T.~A. Maier, H.-B. Cao,
  S.-X. Chi, M.~Matsuda, Y.~Wang, M.~R. Koehler, P.~R.~C. Kent, and H.~H.
  Weitering.
\newblock Itinerant antiferromagnetism in ${\mathrm{ruo}}_{2}$.
\newblock \emph{Phys. Rev. Lett.}, 118:\penalty0 077201, Feb 2017.
\newblock \doi{10.1103/PhysRevLett.118.077201}.
\newblock URL \url{https://link.aps.org/doi/10.1103/PhysRevLett.118.077201}.

\bibitem[Šmejkal et~al.(2020{\natexlab{a}})Šmejkal, González-Hernández,
  Jungwirth, and Sinova]{Smejkal2020}
Libor Šmejkal, Rafael González-Hernández, T.~Jungwirth, and J.~Sinova.
\newblock Crystal time-reversal symmetry breaking and spontaneous hall effect
  in collinear antiferromagnets.
\newblock \emph{Science Advances}, 6\penalty0 (23):\penalty0 eaaz8809,
  2020{\natexlab{a}}.
\newblock \doi{10.1126/sciadv.aaz8809}.
\newblock URL \url{https://www.science.org/doi/abs/10.1126/sciadv.aaz8809}.

\bibitem[Zhou et~al.(2024)Zhou, Feng, Zhang, \ifmmode~\check{S}\else
  \v{S}\fi{}mejkal, Sinova, Mokrousov, and Yao]{Yao2024}
Xiaodong Zhou, Wanxiang Feng, Run-Wu Zhang, Libor \ifmmode~\check{S}\else
  \v{S}\fi{}mejkal, Jairo Sinova, Yuriy Mokrousov, and Yugui Yao.
\newblock Crystal thermal transport in altermagnetic ${\mathrm{ruo}}_{2}$.
\newblock \emph{Phys. Rev. Lett.}, 132:\penalty0 056701, Jan 2024.
\newblock \doi{10.1103/PhysRevLett.132.056701}.
\newblock URL \url{https://link.aps.org/doi/10.1103/PhysRevLett.132.056701}.

\bibitem[Fernandes et~al.(2024)Fernandes, de~Carvalho, Birol, and
  Pereira]{Fernandes2024pintchpoint}
Rafael~M. Fernandes, Vanuildo~S. de~Carvalho, Turan Birol, and Rodrigo~G.
  Pereira.
\newblock Topological transition from nodal to nodeless zeeman splitting in
  altermagnets.
\newblock \emph{Phys. Rev. B}, 109:\penalty0 024404, Jan 2024.
\newblock \doi{10.1103/PhysRevB.109.024404}.
\newblock URL \url{https://link.aps.org/doi/10.1103/PhysRevB.109.024404}.

\bibitem[Antonenko et~al.(2024)Antonenko, Fernandes, and
  Venderbos]{Fernandes2024mirror}
Daniil~S Antonenko, Rafael~M Fernandes, and Jorn~WF Venderbos.
\newblock Mirror chern bands and weyl nodal loops in altermagnets.
\newblock \emph{arXiv preprint arXiv:2402.10201}, 2024.

\bibitem[Bhowal and Spaldin(2024)]{Spaldin2024}
Sayantika Bhowal and Nicola~A. Spaldin.
\newblock Ferroically ordered magnetic octupoles in $d$-wave altermagnets.
\newblock \emph{Phys. Rev. X}, 14:\penalty0 011019, Feb 2024.
\newblock \doi{10.1103/PhysRevX.14.011019}.
\newblock URL \url{https://link.aps.org/doi/10.1103/PhysRevX.14.011019}.

\bibitem[Steward et~al.(2023)Steward, Fernandes, and Schmalian]{Joerg2023}
Charles R.~W. Steward, Rafael~M. Fernandes, and J\"org Schmalian.
\newblock Dynamic paramagnon-polarons in altermagnets.
\newblock \emph{Phys. Rev. B}, 108:\penalty0 144418, Oct 2023.
\newblock \doi{10.1103/PhysRevB.108.144418}.
\newblock URL \url{https://link.aps.org/doi/10.1103/PhysRevB.108.144418}.

\bibitem[McClarty and Rau(2024)]{Rau2024}
Paul~A. McClarty and Jeffrey~G. Rau.
\newblock Landau theory of altermagnetism.
\newblock \emph{Phys. Rev. Lett.}, 132:\penalty0 176702, Apr 2024.
\newblock \doi{10.1103/PhysRevLett.132.176702}.
\newblock URL \url{https://link.aps.org/doi/10.1103/PhysRevLett.132.176702}.

\bibitem[{\=O}ik{\'e} et~al.(2024){\=O}ik{\'e}, Shinada, and
  Peters]{Peters2024}
Jun {\=O}ik{\'e}, Koki Shinada, and Robert Peters.
\newblock Nonlinear magnetoelectric effect under magnetic octupole order: Its
  application to a $ d $-wave altermagnet and a pyrochlore lattice with
  all-in/all-out magnetic order.
\newblock \emph{arXiv preprint arXiv:2407.15836}, 2024.

\bibitem[Fang et~al.(2024)Fang, Cano, and Ghorashi]{Ghorashi2024}
Yuan Fang, Jennifer Cano, and Sayed Ali~Akbar Ghorashi.
\newblock Quantum geometry induced nonlinear transport in altermagnets.
\newblock \emph{Phys. Rev. Lett.}, 133:\penalty0 106701, Sep 2024.
\newblock \doi{10.1103/PhysRevLett.133.106701}.
\newblock URL \url{https://link.aps.org/doi/10.1103/PhysRevLett.133.106701}.

\bibitem[Sorn and Patri(2024)]{Sorn2024}
Sopheak Sorn and Adarsh~S. Patri.
\newblock Signatures of hidden octupolar order from nonlinear hall effects.
\newblock \emph{Phys. Rev. B}, 110:\penalty0 125127, Sep 2024.
\newblock \doi{10.1103/PhysRevB.110.125127}.
\newblock URL \url{https://link.aps.org/doi/10.1103/PhysRevB.110.125127}.

\bibitem[Rao et~al.(1968)Rao, Sherwood, and Bartlett]{Rao1968}
RP~Rao, RC~Sherwood, and N~Bartlett.
\newblock “weak” ferromagnetism in pdf2.
\newblock \emph{The Journal of Chemical Physics}, 49\penalty0 (8):\penalty0
  3728--3730, 1968.

\bibitem[Zhu et~al.(2019)Zhu, Strempfer, Rao, Occhialini, Pelliciari, Choi,
  Kawaguchi, You, Mitchell, Shao-Horn, and Comin]{Comin2019}
Z.~H. Zhu, J.~Strempfer, R.~R. Rao, C.~A. Occhialini, J.~Pelliciari, Y.~Choi,
  T.~Kawaguchi, H.~You, J.~F. Mitchell, Y.~Shao-Horn, and R.~Comin.
\newblock Anomalous antiferromagnetism in metallic ${\mathrm{ruo}}_{2}$
  determined by resonant x-ray scattering.
\newblock \emph{Phys. Rev. Lett.}, 122:\penalty0 017202, Jan 2019.
\newblock \doi{10.1103/PhysRevLett.122.017202}.
\newblock URL \url{https://link.aps.org/doi/10.1103/PhysRevLett.122.017202}.

\bibitem[Šmejkal et~al.(2020{\natexlab{b}})Šmejkal, González-Hernández,
  Jungwirth, and Sinova]{Smejkal2020B}
Libor Šmejkal, Rafael González-Hernández, T.~Jungwirth, and J.~Sinova.
\newblock Crystal time-reversal symmetry breaking and spontaneous hall effect
  in collinear antiferromagnets.
\newblock \emph{Science Advances}, 6\penalty0 (23):\penalty0 eaaz8809,
  2020{\natexlab{b}}.
\newblock \doi{10.1126/sciadv.aaz8809}.
\newblock URL \url{https://www.science.org/doi/abs/10.1126/sciadv.aaz8809}.

\bibitem[Feng et~al.(2022)Feng, Zhou, {\v{S}}mejkal, Wu, Zhu, Guo,
  Gonz{\'a}lez-Hern{\'a}ndez, Wang, Yan, Qin, Zhang, Wu, Chen, Meng, Liu, Xia,
  Sinova, Jungwirth, and Liu]{Feng2022}
Zexin Feng, Xiaorong Zhou, Libor {\v{S}}mejkal, Lei Wu, Zengwei Zhu, Huixin
  Guo, Rafael Gonz{\'a}lez-Hern{\'a}ndez, Xiaoning Wang, Han Yan, Peixin Qin,
  Xin Zhang, Haojiang Wu, Hongyu Chen, Ziang Meng, Li~Liu, Zhengcai Xia, Jairo
  Sinova, Tom{\'a}{\v{s}} Jungwirth, and Zhiqi Liu.
\newblock An anomalous hall effect in altermagnetic ruthenium dioxide.
\newblock \emph{Nature Electronics}, 5\penalty0 (11):\penalty0 735--743, Nov
  2022.
\newblock ISSN 2520-1131.
\newblock \doi{10.1038/s41928-022-00866-z}.
\newblock URL \url{https://doi.org/10.1038/s41928-022-00866-z}.

\bibitem[{\v{S}}mejkal et~al.(2022){\v{S}}mejkal, MacDonald, Sinova, Nakatsuji,
  and Jungwirth]{Smejkal2022}
Libor {\v{S}}mejkal, Allan~H. MacDonald, Jairo Sinova, Satoru Nakatsuji, and
  Tomas Jungwirth.
\newblock Anomalous hall antiferromagnets.
\newblock \emph{Nature Reviews Materials}, 7\penalty0 (6):\penalty0 482--496,
  Jun 2022.
\newblock ISSN 2058-8437.
\newblock \doi{10.1038/s41578-022-00430-3}.
\newblock URL \url{https://doi.org/10.1038/s41578-022-00430-3}.

\bibitem[Wang et~al.(2023)Wang, Tanaka, Sakai, Wang, Deng, Lyu, Li, Tian, Shen,
  Ogawa, Kanazawa, Yu, Arita, and Kagawa]{Wang2023}
Meng Wang, Katsuhiro Tanaka, Shiro Sakai, Ziqian Wang, Ke~Deng, Yingjie Lyu,
  Cong Li, Di~Tian, Shengchun Shen, Naoki Ogawa, Naoya Kanazawa, Pu~Yu, Ryotaro
  Arita, and Fumitaka Kagawa.
\newblock Emergent zero-field anomalous hall effect in a reconstructed rutile
  antiferromagnetic metal.
\newblock \emph{Nature Communications}, 14\penalty0 (1):\penalty0 8240, Dec
  2023.
\newblock ISSN 2041-1723.
\newblock \doi{10.1038/s41467-023-43962-0}.
\newblock URL \url{https://doi.org/10.1038/s41467-023-43962-0}.

\bibitem[Colom\'es and Franz(2018)]{Franz2018}
E.~Colom\'es and M.~Franz.
\newblock Antichiral edge states in a modified haldane nanoribbon.
\newblock \emph{Phys. Rev. Lett.}, 120:\penalty0 086603, Feb 2018.
\newblock \doi{10.1103/PhysRevLett.120.086603}.
\newblock URL \url{https://link.aps.org/doi/10.1103/PhysRevLett.120.086603}.

\bibitem[Denner et~al.(2020)Denner, Lado, and Zilberberg]{Zilberberg2020}
M.~Michael Denner, J.~L. Lado, and Oded Zilberberg.
\newblock Antichiral states in twisted graphene multilayers.
\newblock \emph{Phys. Rev. Res.}, 2:\penalty0 043190, Nov 2020.
\newblock \doi{10.1103/PhysRevResearch.2.043190}.
\newblock URL \url{https://link.aps.org/doi/10.1103/PhysRevResearch.2.043190}.

\bibitem[Medina Due\~nas et~al.(2022)Medina Due\~nas, Calvo, and
  Foa~Torres]{Foa2022}
Joaqu\'{\i}n Medina Due\~nas, Hern\'an~L. Calvo, and Luis E.~F. Foa~Torres.
\newblock Copropagating edge states produced by the interaction between
  electrons and chiral phonons in two-dimensional materials.
\newblock \emph{Phys. Rev. Lett.}, 128:\penalty0 066801, Feb 2022.
\newblock \doi{10.1103/PhysRevLett.128.066801}.
\newblock URL \url{https://link.aps.org/doi/10.1103/PhysRevLett.128.066801}.

\bibitem[Mella et~al.(2023)Mella, Calvo, and Foa~Torres]{Luis2023}
Jos{\'e}~D. Mella, Hernán~L. Calvo, and Luis E.~F. Foa~Torres.
\newblock Entangled states induced by electron–phonon interaction in
  two-dimensional materials.
\newblock \emph{Nano Letters}, 23\penalty0 (23):\penalty0 11013--11018, 2023.
\newblock \doi{10.1021/acs.nanolett.3c03316}.
\newblock URL \url{https://doi.org/10.1021/acs.nanolett.3c03316}.
\newblock PMID: 37984421.

\bibitem[Su and Li(2024)]{Li2024}
Yunlong Su and Gang Li.
\newblock Topological origin of antichiral edge states induced by a nonchiral
  phonon.
\newblock \emph{Phys. Rev. B}, 109:\penalty0 155410, Apr 2024.
\newblock \doi{10.1103/PhysRevB.109.155410}.
\newblock URL \url{https://link.aps.org/doi/10.1103/PhysRevB.109.155410}.

\bibitem[Mazin et~al.(2021)Mazin, Koepernik, Johannes, González-Hernández,
  and Šmejkal]{Mazin2021}
Igor~I. Mazin, Klaus Koepernik, Michelle~D. Johannes, Rafael
  González-Hernández, and Libor Šmejkal.
\newblock Prediction of unconventional magnetism in doped fesb<sub>2</sub>.
\newblock \emph{Proceedings of the National Academy of Sciences}, 118\penalty0
  (42):\penalty0 e2108924118, 2021.
\newblock \doi{10.1073/pnas.2108924118}.
\newblock URL \url{https://www.pnas.org/doi/abs/10.1073/pnas.2108924118}.

\bibitem[Su et~al.(1979)Su, Schrieffer, and Heeger]{SSH}
W.~P. Su, J.~R. Schrieffer, and A.~J. Heeger.
\newblock Solitons in polyacetylene.
\newblock \emph{Phys. Rev. Lett.}, 42:\penalty0 1698--1701, Jun 1979.
\newblock \doi{10.1103/PhysRevLett.42.1698}.
\newblock URL \url{https://link.aps.org/doi/10.1103/PhysRevLett.42.1698}.

\bibitem[Cano et~al.(2018)Cano, Bradlyn, Wang, Elcoro, Vergniory, Felser,
  Aroyo, and Bernevig]{Cano2018}
Jennifer Cano, Barry Bradlyn, Zhijun Wang, L.~Elcoro, M.~G. Vergniory,
  C.~Felser, M.~I. Aroyo, and B.~Andrei Bernevig.
\newblock Building blocks of topological quantum chemistry: Elementary band
  representations.
\newblock \emph{Phys. Rev. B}, 97:\penalty0 035139, Jan 2018.
\newblock \doi{10.1103/PhysRevB.97.035139}.
\newblock URL \url{https://link.aps.org/doi/10.1103/PhysRevB.97.035139}.

\bibitem[Po et~al.(2018)Po, Watanabe, and Vishwanath]{Vishwanath2018}
Hoi~Chun Po, Haruki Watanabe, and Ashvin Vishwanath.
\newblock Fragile topology and wannier obstructions.
\newblock \emph{Phys. Rev. Lett.}, 121:\penalty0 126402, Sep 2018.
\newblock \doi{10.1103/PhysRevLett.121.126402}.
\newblock URL \url{https://link.aps.org/doi/10.1103/PhysRevLett.121.126402}.

\bibitem[Khalaf et~al.(2021)Khalaf, Benalcazar, Hughes, and
  Queiroz]{Queiroz2021}
Eslam Khalaf, Wladimir~A. Benalcazar, Taylor~L. Hughes, and Raquel Queiroz.
\newblock Boundary-obstructed topological phases.
\newblock \emph{Phys. Rev. Res.}, 3:\penalty0 013239, Mar 2021.
\newblock \doi{10.1103/PhysRevResearch.3.013239}.
\newblock URL \url{https://link.aps.org/doi/10.1103/PhysRevResearch.3.013239}.

\bibitem[Xu et~al.(2024)Xu, Elcoro, Song, Vergniory, Felser, Parkin, Regnault,
  Ma\~nes, and Bernevig]{Bernevig2024}
Yuanfeng Xu, Luis Elcoro, Zhi-Da Song, M.~G. Vergniory, Claudia Felser, Stuart
  S.~P. Parkin, Nicolas Regnault, Juan~L. Ma\~nes, and B.~Andrei Bernevig.
\newblock Filling-enforced obstructed atomic insulators.
\newblock \emph{Phys. Rev. B}, 109:\penalty0 165139, Apr 2024.
\newblock \doi{10.1103/PhysRevB.109.165139}.
\newblock URL \url{https://link.aps.org/doi/10.1103/PhysRevB.109.165139}.

\bibitem[Kruthoff et~al.(2017)Kruthoff, de~Boer, van Wezel, Kane, and
  Slager]{Slager2017}
Jorrit Kruthoff, Jan de~Boer, Jasper van Wezel, Charles~L. Kane, and Robert-Jan
  Slager.
\newblock Topological classification of crystalline insulators through band
  structure combinatorics.
\newblock \emph{Phys. Rev. X}, 7:\penalty0 041069, Dec 2017.
\newblock \doi{10.1103/PhysRevX.7.041069}.
\newblock URL \url{https://link.aps.org/doi/10.1103/PhysRevX.7.041069}.

\bibitem[Rhim et~al.(2017)Rhim, Behrends, and Bardarson]{Bardarson2017}
Jun-Won Rhim, Jan Behrends, and Jens~H. Bardarson.
\newblock Bulk-boundary correspondence from the intercellular zak phase.
\newblock \emph{Phys. Rev. B}, 95:\penalty0 035421, Jan 2017.
\newblock \doi{10.1103/PhysRevB.95.035421}.
\newblock URL \url{https://link.aps.org/doi/10.1103/PhysRevB.95.035421}.

\bibitem[Bouhon et~al.(2019)Bouhon, Black-Schaffer, and Slager]{Slager2019}
Adrien Bouhon, Annica~M. Black-Schaffer, and Robert-Jan Slager.
\newblock Wilson loop approach to fragile topology of split elementary band
  representations and topological crystalline insulators with time-reversal
  symmetry.
\newblock \emph{Phys. Rev. B}, 100:\penalty0 195135, Nov 2019.
\newblock \doi{10.1103/PhysRevB.100.195135}.
\newblock URL \url{https://link.aps.org/doi/10.1103/PhysRevB.100.195135}.

\bibitem[Delplace et~al.(2011)Delplace, Ullmo, and Montambaux]{Montambaux2011}
P.~Delplace, D.~Ullmo, and G.~Montambaux.
\newblock Zak phase and the existence of edge states in graphene.
\newblock \emph{Phys. Rev. B}, 84:\penalty0 195452, Nov 2011.
\newblock \doi{10.1103/PhysRevB.84.195452}.
\newblock URL \url{https://link.aps.org/doi/10.1103/PhysRevB.84.195452}.

\bibitem[Shen(2012)]{Shen2012}
Shun-Qing Shen.
\newblock \emph{Topological insulators}, volume 174.
\newblock Springer, 2012.

\end{thebibliography}
\end{document}